\documentclass[aps,showkeys,twocolumn]{revtex4}

\usepackage{amsmath,amsfonts,amssymb}
\usepackage{graphicx,graphics,natbib,dcolumn,bm,enumerate}

\begin{document}

\title{Universality in voting behavior: an empirical analysis}

\author{Arnab Chatterjee}
\affiliation{Department of Biomedical Engineering and Computational Science, 
Aalto University School of Science, P.O.  Box 12200, FI-00076, Finland} 
\author{Marija Mitrovi\'{c}}
\affiliation{Department of Biomedical Engineering and Computational Science, 
Aalto University School of Science, P.O.  Box 12200, FI-00076, Finland} 
\author{Santo Fortunato}
\email[Email: ]{santo.fortunato@aalto.fi}
\affiliation{Department of Biomedical Engineering and Computational Science, 
Aalto University School of Science, P.O.  Box 12200, FI-00076, Finland}

\begin{abstract}
Election data represent a precious source of information to study
human behavior at a large scale. In proportional elections with open
lists, the number of votes received by a candidate, rescaled
by the average performance of all competitors in the same party list, 
has the same
distribution regardless of the country and the year of the election.
Here we provide the first thorough assessment of this claim.
We analyzed election datasets of 15 countries with proportional
systems. We confirm that a class of nations with similar election
rules fulfill the universality claim. Discrepancies from this trend
in other countries with open-lists elections are always associated with 
peculiar differences in the election rules, which matter
more than differences between countries and historical periods.
Our analysis shows that the role of parties in the electoral
performance of candidates is crucial: alternative scalings not
taking into account party affiliations lead to poor results.
\end{abstract}

\maketitle

\section{Introduction}
We know from statistical physics that systems of many particles
exhibit, in the aggregate, a behavior which is enforced by a few basic
features of the individual particles, but independent of all other
characteristics. This result is particularly striking in critical phenomena,
like continuous phase transitions and is known as {\it universality}~\cite{binney92}.
Empirical evidence shows that a number of social phenomena are also
characterized by simple emergent behavior out of the interactions of
many individuals. The most striking example is collective motion
\cite{helbing01,helbing02,vicsek12}. Therefore, in the last years a
growing community of scholars have been
analyzing large-scale social dynamics and proposing simple microscopic
models to describe it, alike the minimalistic models used in statistical physics. 
Such scientific endeavour, initially known by the name of {\it sociophysics}~\cite{ball04,buchanan07,castellano09}, has been
meanwhile augmented by scholars and tools of other disciplines, like
applied mathematics, social and computer science, and is currently referred to as {\it computational social science}~\cite{lazer09}. 

Elections are among the largest social phenomena. 
In India, USA and Brazil hundreds of million voters cast their preferences on
election day. Fortunately,
datasets can be freely downloaded from institutional sources,
like the Ministry of Internal Affairs of many countries. Therefore, it
is not surprising that elections have been among the most studied
social phenomena of the last decade~\cite{fortunato12a}. By now, several aspects of voting
behavior have been examined, like statistics of turnout
rates~\cite{borghesi10,borghesi12}, detection of election
anomalies~\cite{baez06,klimek12}, polarization and tactical voting in
mayoral elections~\cite{araripe06,araujo10}, the relation between party size and temporal
correlations~\cite{andresen07}, the relation between 
number of candidates and number of voters~\cite{mantovani11}, the
emergence of third parties in bipartisan systems~\cite{romero11}, the correlation between the score of a party and
the number of its members~\cite{schneider05b}, the classification of
electoral campaigns~\cite{sadovsky07}, etc.

The most studied feature is the distribution of
the number of votes of
candidates~\cite{costafilho99,bernardes02,costafilho03,lyra03,gonzalez04,
situngkir04,sousa05,morales06,travieso06,fortunato07,gradowski08,araripe09,hernandez-saldana09,chou09,banisch10}.
In the first analysis by Costa Filho et al.~\cite{costafilho99}, the
distribution of the fraction of votes received by candidates in
Brazilian federal and state elections seems to
decay as a power law with exponent $-1$ in the central
region. Following this finding several similar analyses have been
performed on election data of various countries, like
India~\cite{gonzalez04}, Indonesia~\cite{situngkir04} and Mexico~\cite{morales06}.

However, Fortunato and Castellano observed that the analysis by
Costa Filho et al. treats all candidates equally, neglecting the role
of the party in the electoral performance~\cite{fortunato07}. This
assumptions appears too strong and unjustified, as the final score of
the candidate is likely to depend on whether his/her party is popular
or not. For this reason Fortunato and Castellano argued that
characterizing and modelling
the competition of candidates of the same party is more promising, as
the performance of the candidates would be mostly depending on
their own activity, rather than on external factors. Such competition
occurs in a peculiar type of voting system, viz. proportional
elections with open lists, where people may vote for a party and one
or more candidates. In this system, people may actually choose their
representatives by voting directly for them, whereas the number of
candidates entering the Parliament for a given party typically depends
on the strength of the party at the national and/or regional level.
In these elections, it was found that the distributions of the number
of votes of a candidate, divided by the average number of votes of
all party competitors in the same list, appear to be the same
regardless of the country and the year of the election~\cite{fortunato07}. This claim has
been recently disputed by Araripe and Costa Filho, who found that the
universal curve computed in Ref.~\cite{fortunato07} does not
follow well the profile of the distribution of Brazilian elections,
which are also proportional and with open lists. 

Here we carry out the first comprehensive analysis of the distribution
of candidates' performance, using election results of 15
countries. We focus on proportional elections, as they feature
the open-list system that allow voters to choose their
representatives, enabling a real competition between candidates. 
We conclude that the relative performance,
i.e. the ratio between the number of votes of a candidate and the
average score of his/her party competitors in a given list has 
indeed the same distribution for countries with similar voting systems, and
that discrepancies from the universal distribution emerge when the
election has markedly different features (e.g. large districts,
compulsory voting and weak role of parties in Brazil).
We also show that party affiliations cannot be neglected:
statistics of the absolute performance of candidates of different parties,
like that investigated in the original
analysis by Costa Filho et al., do not compare well between countries. 

\section{Results}
\subsection{Proportional elections}
The electoral system we wish to study
is proportional representation (PR)~\cite{Villodres06}. 
We analyze data from parliamentary elections of $15$ 
countries: Italy (before 1992), Poland, Finland,  Denmark, Estonia,
Sweden, Belgium, Switzerland, Slovenia, Czech Republic, Greece, Slovakia, 
Netherlands, Uruguay and Brazil. 
The basic principle 
is that all voters deserve representation and all political groups 
deserve to be represented in legislatures in proportion to their 
strength in the electorate. 
In order to achieve this `fair' representation, the country is usually 
divided into multi-member districts, each district in turn allocating a
certain number of seats. Most countries having a PR system use a party
list voting scheme to allocate the seats among the parties -- each political party
presents a list of candidates for each district. 
On the ballot the voters indicate their preference to a political
party by selecting one or more candidates from the list. 
The number of seats assigned to each party in a district is proportional to the
number of votes. 
The party list systems can be categorized into \textit{open}, \textit{semi-open} 
and \textit{closed}. 

\paragraph*{Open lists}
Open lists enable voters to express their preference not only among
parties but also among candidates. A party presents an unordered, random or
alphabetically ordered list of candidates. Voters choose one or
more candidates, and not the party. The position of each candidate depends
entirely on the number of votes that he/she receives.
In this category, we have studied data from 
Italy (before 1994, when a new system was introduced), Poland, Finland, Denmark, Estonia (since 2002), Greece, 
Switzerland, Slovenia, Brazil, Uruguay.

\paragraph*{Semi-open lists}
Semi-open lists impose some restrictions on voters directly or indirectly.
The voter votes for either a party or a candidate within 
a party list. The parties usually put up a list of candidates according
to their `initial' preference, which depends on internal party
rankings, etc. Candidates conquer parliamentary seats in the order they
are ranked in the list, from the first to the last. However, if a
candidate gets a number of votes exceeding a threshold, then he/she
climbs up the ranking even if he/she was initially at the bottom of the list. 
The final order of the candidates
is decided based on the `initial' ordering
and the actual votes received by the candidates.
Sweden, Slovakia, Czech Republic, Belgium, Estonia (until 2002) and Netherlands
fall in this category.

\paragraph*{Closed lists}
In the closed list system the party fixes the order in which the candidates are
listed and elected. The voter casts a vote to a party as a whole and cannot
express his/her preference for any candidate or group of candidates. 
The representatives are then selected as they appear on the list,
in the order defined before the elections. Countries voting with this
system include Russia, Italy (since 2006), Spain, Angola, South Africa, Israel, Sri
Lanka, Hong Kong, Argentina, etc. We did not
consider this type of elections in our analysis, as there is no real
competition between the candidates.

The allocation of seats to the parties takes place according
to some pre-defined method, 
e.g. \textit{d'Hondt}, \textit{Hagenbach-Bischoff},  \textit{Sainte-Lagu\"{e}}, 
or some modified version of these~\cite{colomer04}.

\subsection{Distribution of candidates' performance: open lists}
\begin{figure*}[tb]
  \centering
  {\includegraphics[width=0.75\textwidth]{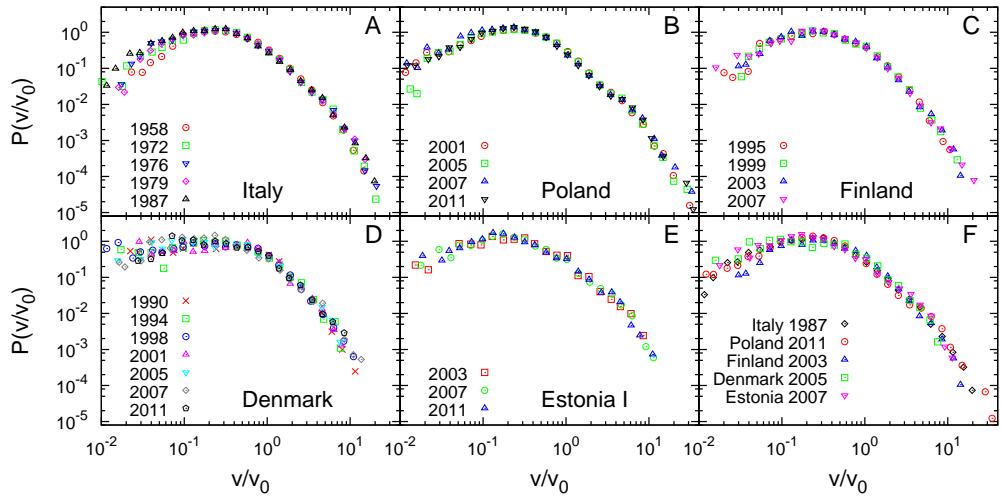}}
\caption{Distribution of electoral
  performance of candidates in proportional elections with open
  lists, according to FC scaling. Italy (until 1992), Poland, Finland,
Denmark and Estonia (after 2002) follow essentially the same rules,
which is reflected by the data collapse of panel F. The historical
evolution of the countries does not seem to affect the shape of the
distribution (panels A to E). }
\label{fig1}
\end{figure*}
\begin{figure*}[tb]
  \centering
  {\includegraphics[width=0.75\textwidth]{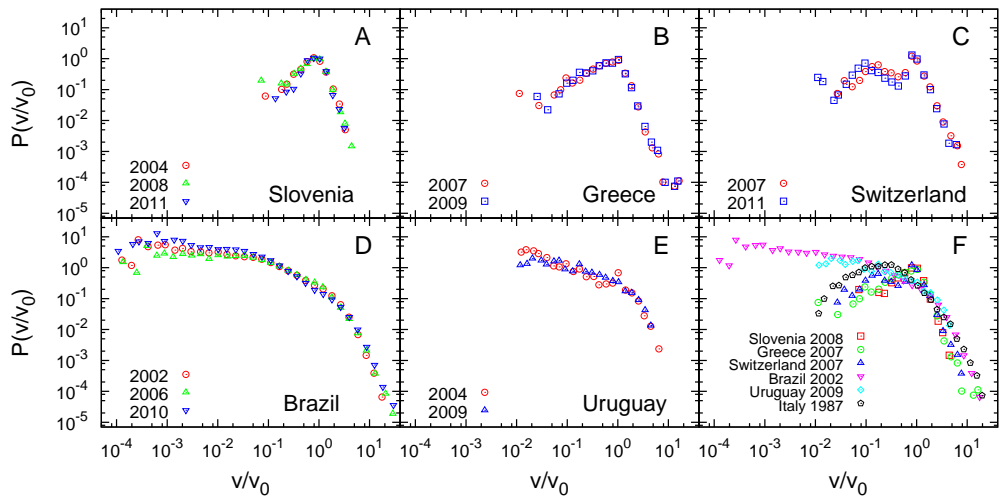}}
\caption{Same analysis as in Fig. 1, for Slovenia, Greece,
  Switzerland, Brazil and Uruguay. Curves are
  fairly stable at the national level, but they do not compare well
  across countries and with the universal curves of Fig.1 (represented
  in panel F by the distribution for the Italian elections in
  1987). Such discrepancies are likely to be due to the different 
  election rules of these countries as compared to each other and to the ones examined in
Fig. 1, although they all adopt open lists.}
\label{fig2}
\end{figure*}
In every proportional election, the country is divided into districts and each party
presents a list with $Q$ candidates. Voters typically choose one of the parties and 
express their preference among the candidates of the selected party. The seat
allocation depends on the country (see Section~\ref{app1} of Appendix) 
and has a large influence on how voters choose who they will vote for. 
The data sets we considered contain information about the number of votes $v_{i}$
that each candidate $i$ received and the number of candidates $Q_{i}$ of
the party list $l_{i}$ including candidate $i$. From this information
one can derive the total number of votes $N_{l_i}$ collected by the $Q_{i}$
candidates of list $l_{i}$. By summing over all party lists in the
district $D^i$ of candidate $i$ we obtain the number of votes
$N_{D^i}$ in the district. The total number of votes cast during the
whole election is indicated as $N_T$.

Our analysis consists in computing the probability distribution of the
number of votes received by candidates, suitably normalized. We use the
following normalizations:
\begin{itemize}
\item{The scaling by Fortunato and Castellano~\cite{fortunato07},
    where the number of votes $v^i$ of a candidate is divided by the average
  number of votes $v_0=N_{l_i}/Q^i$ of all candidates in his/her party
  list. We shall indicate it as FC scaling.}
\item{The scaling by Costa Filho, Almeida, Andrade and Moreira (CAAM)~\cite{costafilho99}, where one
  considers the fraction of votes received by a candidate. Since it
  is unclear to us what one exactly means by that, we consider two possible
normalizations: a) the fraction of the total votes in the district,
$v^i/N_{D^i}$; b) the fraction of the total votes in the country
$v^i/N_T$. We shall refer to them as to CAAMd and CAAMn,
respectively. We rule out the fraction of votes in the party list
because the authors made clear that they do not consider party
affiliations. The most sensible definition appears the normalization
at the district level, which will be thus reported here. The results
for CAAMn are shown in the Appendix (Figs.~\ref{figS1}, \ref{figS2}, \ref{figS3}).}
\end{itemize} 
The universality discovered in Ref.~\cite{fortunato07} referred to
elections held in Finland, Poland and Italy in various years. 
Here we confirm the result with a larger number of
data sets (Fig. \ref{fig1}). Panels A, B and C display the
distributions for Italy, Poland and Finland, respectively, in
different years. The stability of the curve within the same country is
remarkable, especially on the tail. In panel F we compare the
distributions across the countries, yielding the collapse found in
Ref.~\cite{fortunato07}. Elections data in Denmark and Estonia
(detailed in panels D and E), appear to follow the universal curve as well.
We indicate this class of countries as Group U in the following.
In Ref.~\cite{fortunato07} it was shown that this universal curve
is very well represented by a log-normal function.
%
%\begin{equation}
% P(v/v_{0})=\frac{v_{0}}{\sqrt{2\pi}\sigma
%v}\exp \left[-\frac{(\log(v/v_{0})-\mu)^{2}}{2\sigma^{2}}\right],
%\label{FC:ln}
%\end{equation}
%
%with $\mu=-0.54$, $\sigma^{2}=-2\mu=1.08$.

\begin{figure*}[t]
  \centering
  \includegraphics[width=0.75\textwidth]{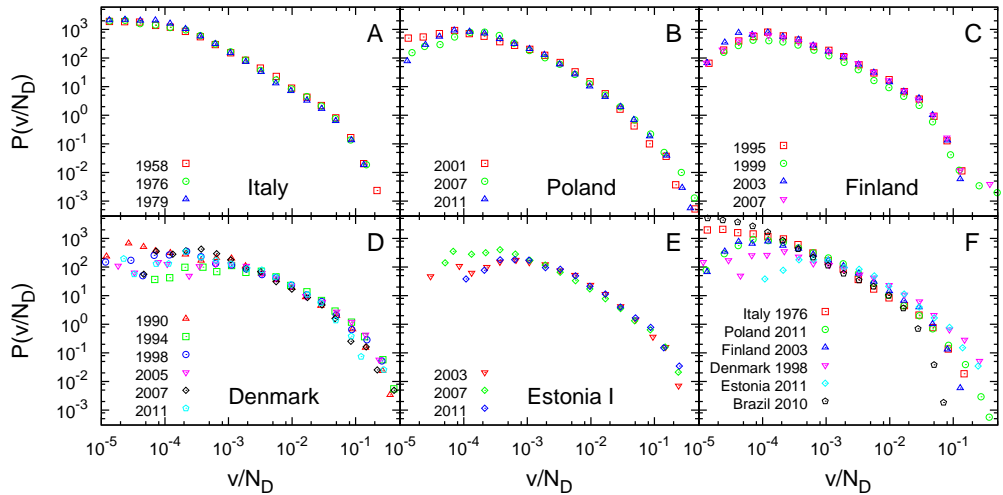}
\caption{Same analysis as in Fig. 1, with CAAMd scaling. Curves are
  stable at the national level, but they do not compare well across countries.}
\label{fig3}
\end{figure*}
\begin{figure*}[t]
  \centering
  {\includegraphics[width=0.75\textwidth]{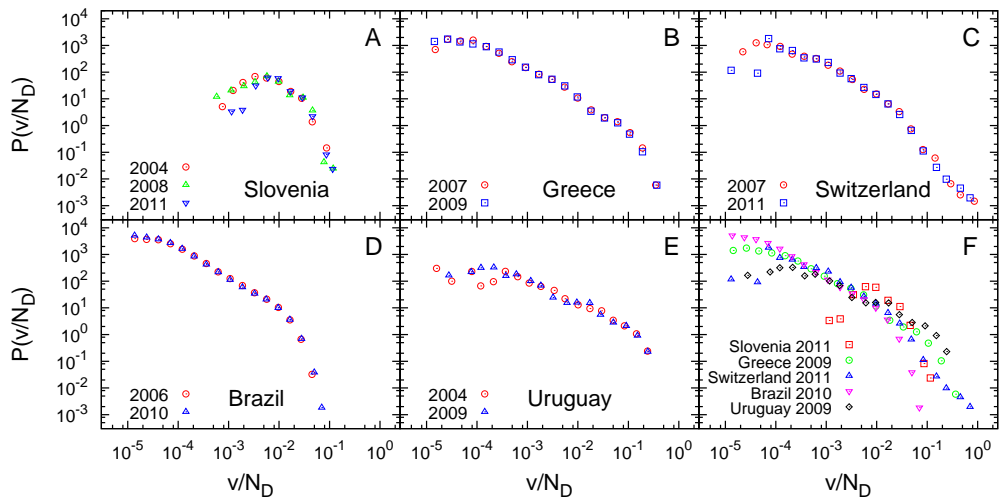}}
\caption{Same analysis as in Fig. 2, with CAAMd scaling. Curves are
  stable at the national level, but they do not compare well across countries.}
\label{fig4}
\end{figure*}

Italy (until 1992), Poland, Finland, Denmark and Estonia (after 2002) 
use open lists~\cite{Villodres06}, in which 
voters can express their preference toward certain candidates within the party
list and have a direct influence on the list ordering. 
These lists use the plurality method for the
allocation of the seats within the party lists: candidates with the largest
number of nominative votes are declared elected. There are just small
differences in the number of candidates that a voter can indicate, the
ordering of the candidates on the ballot, but the systems are
basically the same, justifying the observed universality.

Other countries using open lists are Slovenia, Greece, Switzerland,
Brazil, Uruguay. The results of the FC scaling are illustrated in
Fig~\ref{fig2}. While there is a historical persistence of the
distribution at the national level, the curves do not really follow a
common pattern, and do not match well the behavior of the universal
distribution found for Italy, Poland, Finland, Denmark and
Estonia. We distinguish here two classes of behaviors: 
Slovenia, Greece and Switzerland are characterized by a pronounced
peak at $v/v_0=1$, and their tails match each other quite well. Brazil
and Uruguay exhibit a monotonic pattern, quite different from the other three curves. The Brazilian curve
follows quite closely the profile of the universal curve of Fig. 1 on the tail ($v/v_0>1$).

We conclude that open
list systems do not guarantee identical distributions, but can be
grouped in classes of behaviors. A close inspection of the election
systems, however, may explain why we observe discrepancies. Slovenia
divides its territory into eight districts which in turn 
are partitioned into $11$ electoral units, each giving one candidate in 
the district. The voters can cast the vote for any of the candidates in
the district, but the election of the candidate depends on the number of votes
he/she won in his/her unit, i.e. the performance of the candidate in the unit is
more important than the number of votes won in the district, which may
affect both the candidates' campaigns and the voters' choices.

Greece uses a very complex seat allocation method among
party lists and individual candidates. Although the ranking of the candidates
on the list and the seats reallocation depends on the number of votes collected by the 
candidate, if one of the candidates happens to be the head of a party or a current or ex Prime
Minister he/she is set automatically at the top of the party list,
regardless of his/her electoral performance. Additionally, voting is
compulsory, so many people cast a vote because they have to, without
an informed opinion and/or motivation to participate in the election.

In Switzerland, voters may cast as many votes as there are
seats in the district. They may vote for all members of the
list, or for candidates of more than one party. Voters are
also allowed to cast two votes per candidate. 
This type of list is classified as \textit{free list}.

In Brazil, like in Greece, voting is compulsory, and we
cannot exclude that this plays a role on the shape of the
distribution. In addition, each state is just one district, which then
comprises a number of voters orders of magnitude larger than the
typical districts in all other elections. This explains why the
Brazilian curve spans a much larger range of values for the
performance variable than all other curves. The
huge number of voters in the same district also explains why parties
present very long lists of candidates (often with over one hundred
names). Finally, the role of parties is very weak; the political
constellation frequently changes, with new parties being created and
old ones being reshaped. 

In Uruguay voters cannot choose candidates, but lists of candidates
presented by the parties, the so-called sub-lemas. Therefore our
analysis focuses on the distribution of performance of sub-lemas,
instead of that of single candidates. 

Figs.~\ref{fig3} and \ref{fig4} show the analogues of Figs. 1 and 2
obtained by using CAAMd
scaling. The historical stability of the corresponding distributions at the
national level holds, however the comparison across countries is poor:
curves appear to cross, not to collapse (panel F). According to Costa
Filho et al.~\cite{costafilho99} the central part of the Brazilian curve follows
a power law, with exponent close to $1$; power law fits of the central
region of the other distributions yield exponents sensibly different
from each other, 
which confirms the crossing of the curves (see Table~\ref{tab:exp} of Appendix). In particular, we
cound not identify any portion of the Polish curve resembling a power law. 
We conclude that
the fraction of votes $v/N_{D}$ collected by a candidate in his/her
electoral district does not follow the same probability distribution
in different countries, not even when they have essentially
identical voting schemes, as in Figs.~\ref{fig1} and ~\ref{fig3}.
\begin{figure*}[tb]
  \centering
  {\includegraphics[width=0.75\textwidth]{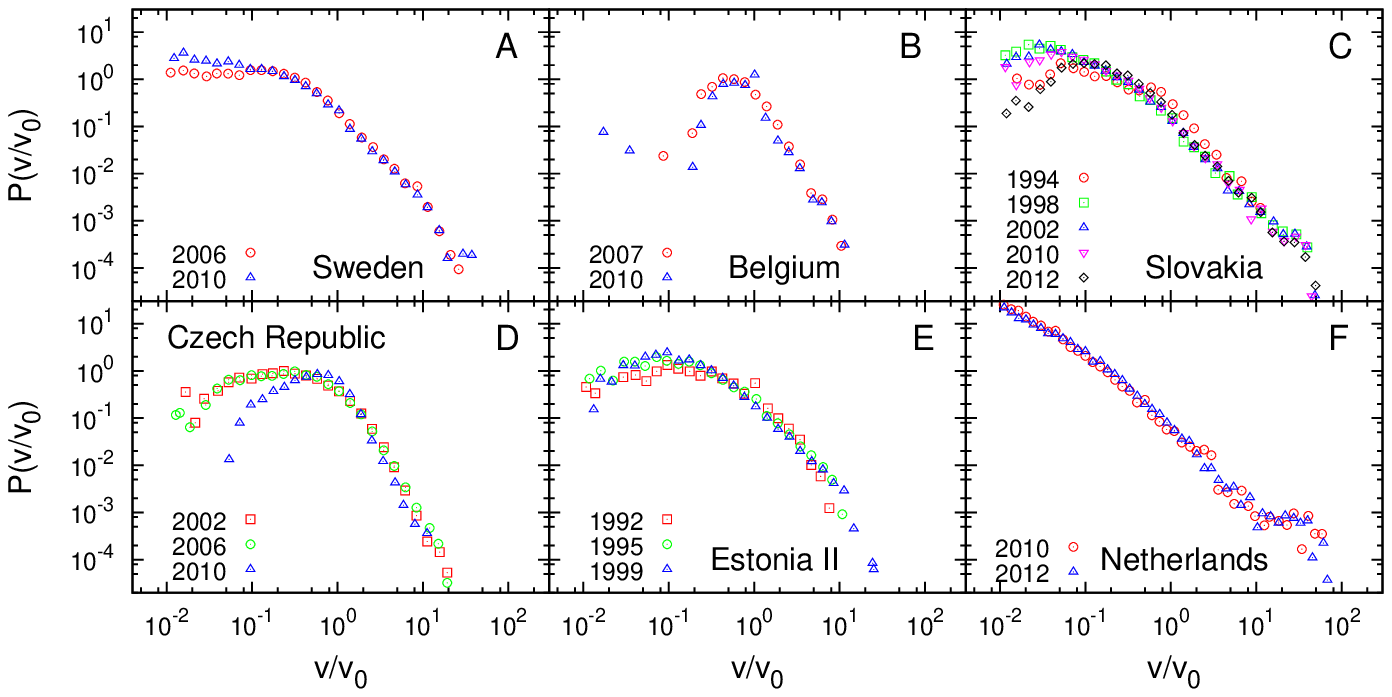}}
\caption{Distribution of electoral
  performance of candidates in proportional elections with semi-open
  lists, according to FC scaling. Voters may express preferences for
  the candidates, but this plays a role for the final seat assignments
  only if the number of votes obtained by a candidate exceeds a given
  threshold, which varies from a country to another. At the national
  level curves are mostly stable, significant discrepancies correspond
to changes in the election rules, like in Slovakia (C), Czech Republic
(D) and Estonia (E). The apparent power law of the Dutch curve (F) might
be generated by a rich-gets-richer mechanism, since the threshold is
very high (10\% at the national level) and voters typically tend to support the candidates based on
their popularity. We stress that Estonia since 2002 has adopted open
lists, which is why distributions of Estonian elections after 2002 are
illustrated in Figs. 1 and 2 (labeled as Estonia I).}
\label{fig5}
\end{figure*}
\begin{figure*}[tb]
  \centering
  {\includegraphics[width=0.75\textwidth]{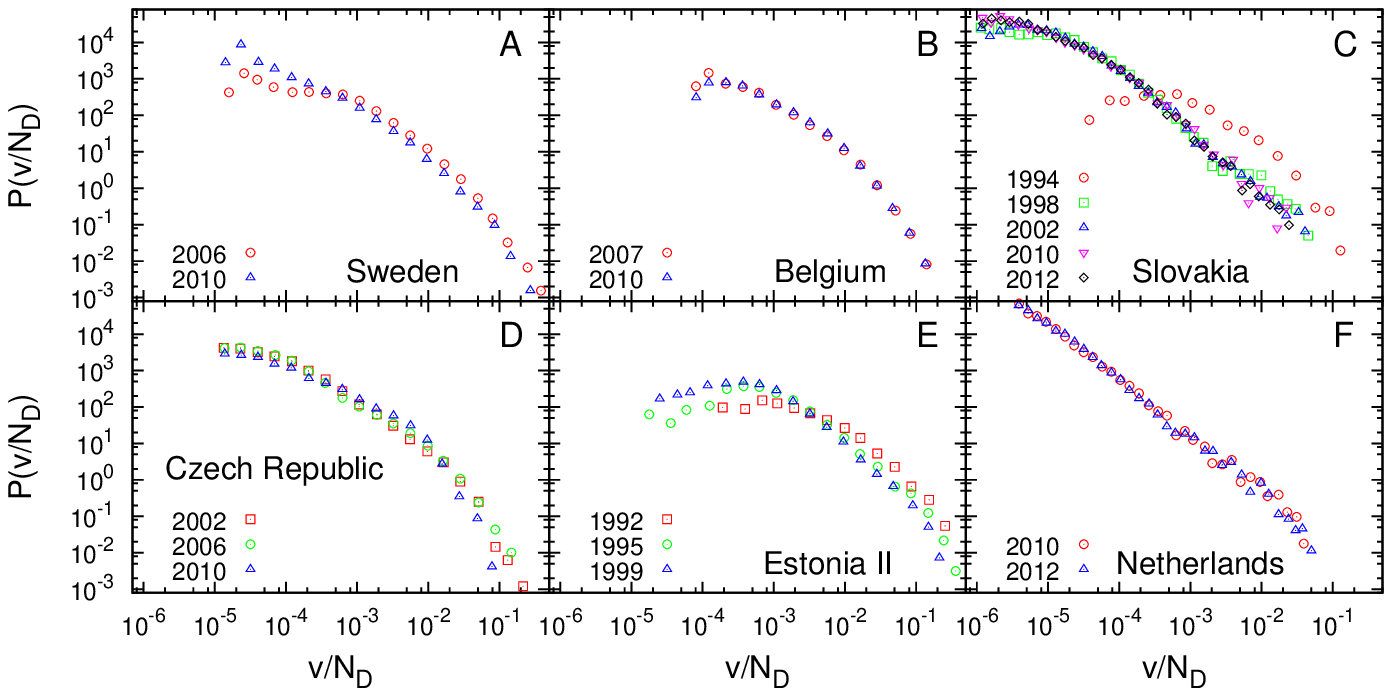}}
\caption{Same as Fig. 5, with CAAMd scaling.}
\label{fig6}
\end{figure*}

\subsection{Distribution of candidates' performance: semi-open lists}
The other countries we considered use semi-open lists, with
different thresholds for the number of preferences that candidates are required to collect in order
to secure a seat in the Parliament. The higher the electoral quota is,
the harder is for a candidate to reach the required number of
votes. In this case the position of the candidate within the party, as
it appears on the ballot, has more influence on his/her final
rank than the number of votes he/she collected. 
This can drastically effect the motivation of the candidate to
lead a personal campaign. Also, high quotas diminish the influence
of the voter on the final list ordering, which affects both the degree
of a candidate's
involvement in his/her personal campaign
and the way people cast their preference votes. Therefore there is hardly an 
open competition between candidates, and this may be reflected in the
shape of the distribution of performance.
Figure~\ref{fig5} shows the probability density distributions
for different countries with semi-open lists, according to FC scaling. The elections in Czech Republic
held in $2010$ had the lowest electoral quota and $P(v/v_{0})$
(Fig.~5D) turns out to be very similar to the curve obtained for Greek elections
(Fig.~3B). The country with the
highest electoral quota are the Netherlands, where each candidate has to win $10\%$ of
votes cast on the national level in order to be directly elected. Voters in
Netherlands have little or no influence on the ordering of 
candidates, which is essentially frozen by the party, and they
often vote for the top-ranked candidate and the first several
names on the list, as they are the most popular and
appreciated members of the party. This resembles the
\textit{rich-gets-richer} effect, which is characterized by
power-law behavior of the distribution of the relevant 
quantities~\cite{Eggenberger23,Simon55,Merton68,Price76,
Albert02}. Indeed, the distribution of performance of Dutch candidates 
follow an approximate power-law behavior over most of the range of the performance
$v/v_{0}$ (Fig.~5F).

Besides the values of the electoral
threshold, these countries also differ in the number of nominative preferences
a voter can cast, in the size and number of multi-member districts, as
well as in the electoral formula that determines the final rankings 
(see Section~\ref{app1} of Appendix). Any change in the
electoral system, i.e. these several factors, might influence the shape of $P(v/v_{0})$. For instance, in $1994$ 
Slovakia changed the number of
multi-member districts, leading to appreciable changes in the
shape of the distribution (Fig.~5C). The change in the
electoral quota and the number of nominative votes decided in Czech
Republic in $2006$, may be the responsible for the variation of the
curve before and after that year
(Fig.~5D). The transition from semi-open to open
lists introduced in Estonia in $2002$, might explain why
the curves before and after that year look different (Fig.~1E versus
Fig.~5E). Interestingly, after the introduction of open lists in
Estonia, the distribution of performance matches the universal
distribution of the other countries with similar election systems (Fig.~1F),
while before $2002$ we find clear discrepancies. 

The corresponding
distributions with CAAMd scaling also show marked differences between
different countries (Fig.~6).

\subsection{Estimating the similarity of the distributions}
\begin{figure*}[t]
  \centering
  {\includegraphics[width=0.75\textwidth]{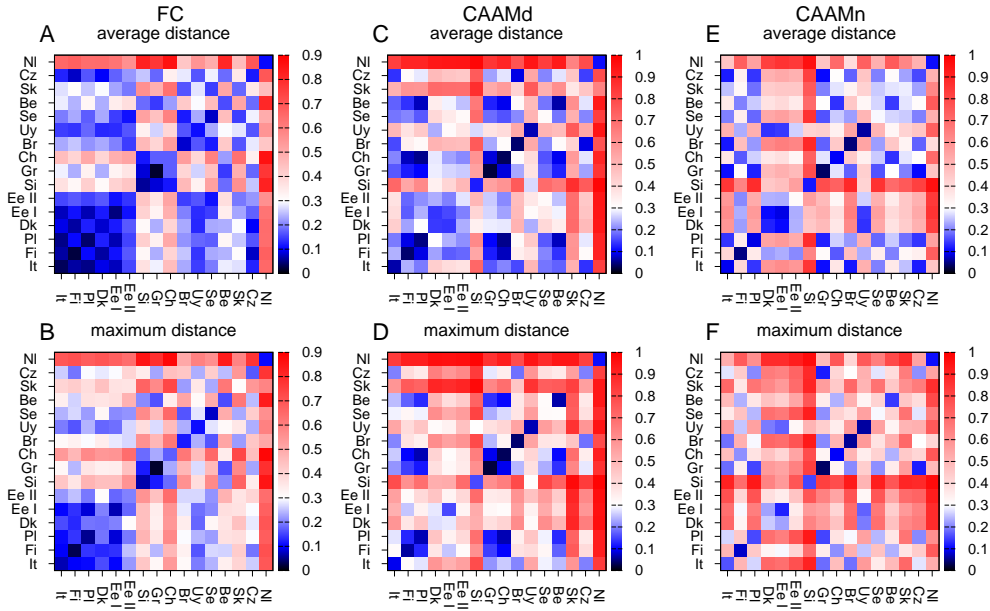}}
\caption{Quantitative assessment of the similarity between
  distributions at the national level and between countries. The
  matrices in the top row indicate the values of the average K-S
  distance between datasets of different countries. On the bottom row
  the maximum distances are reported. Each column corresponds to one
  of the three types of distributions we have examined, by using FC, CAAMd
  and CAAMn scaling. A color code is
  adopted to better distinguish the low values of the distance
  (indicated by the blue), indicating a big similarity between the
  curves, from the larger values, corresponding to poor collapses. The
bluish square on the bottom left of the matrices obtained via FC
scaling confirm that the distributions of those countries are pretty close to each other,
as illustrated in Fig. 1F. Conversely, the similarity between
distributions obtained via CAAMd and CAAMn scaling appears rather
modest for most pairs of countries.}
\label{fig7}
\end{figure*}

So far the estimation of the agreement or disagreement of different
curves has been basically visual. In this section we would like to
attempt a quantitative assessment of this issue. We 
build two matrices, whose entries are the values of the average
distance $D_{avg}$ and the maximum distance
$D_{max}$ between the distributions for any pair of countries for which we gathered election
data (see Methods). 
The dissimilarity values for elections in the same country are
reported on one diagonal of the matrix. Since we have adopted three
different types of scaling for the electoral performance of
candidates, FC, CAAMd and CAAMn, we end up with six matrices, which
are illustrated in Fig.~7. In each column we display the pair of
matrices corresponding to one type of scaling, the first row contains
the average distances, the second row the maximum distances. 
We built $16\times 16$ matrices, even if we studied 15 countries,
because we considered two sets of elections for Estonia, because of
their transition from semi-open lists (Ee II) to open lists (Ee II),
which took place in 2002.

Potential
data collapses are indicated by low values of $D_{avg}$ and $D_{max}$,
which are easier to spot by using a color code, as we did in the
figure. 
Numerical values are listed in the Appendix in the Tables 
\ref{tab:FCclassIa},
\ref{tab:FCclassIm},
\ref{tab:da},
\ref{tab:dm},
\ref{tab:na},
\ref{tab:nm}. 
Dark squares (black-blue) correspond to the lowest values of $D_{avg}$ and
$D_{max}$, so to very similar distributions. The data collapse for the
countries of Group U
(Fig.~1F) is illustrated by the bottom left block of A and
B. Interestingly, we see that only the Estonian elections held after
2002 (Ee I) are very similar to the other curves of Group U; before
2002 Estonians used semi-open lists, the corresponding curves do
not match well with the universal distribution. 

We see
that also the Brazilian and the Uruguayan distributions are fairly
similar, on average, to the universal curve, mostly on the tail,
although they considerably differ in the initial part, especially the
Brazilian distributions. The strong similarity between the results of
elections in the nations of Group U persists even
if we consider the maximum distance (panel B), as the dark block is
still there, though blurred. 
Slovenia and Greece appear very similar to each other but sensibly
different from the other countries.  
The diagonal from bottom left to top right shows the values of the
distance for datasets in the same country. In general, the distances
are pretty low, but we also find fairly large values. These correspond
to countries which introduced changes in the election rules,
reflected in the shape of the distributions, as
described above. 

If we move to CAAMd scaling (panels C and D) the scenario
is considerably worse, in that the curves are much more dissimilar to
each other than the ones obtained with FC scaling. In panel C,
the average distance between the countries of Group U is still
low, though higher than for FC scaling (panel A), but when one moves
to the maximum distance the block disappears (panel D). For CAAMn
(panels E and F) the curves are even more dissimilar to each other.

We are not giving here any indication on the
significance of the measured values of the K-S distance. 
Large values indicate with certainty that the corresponding
distributions are really different curves, but low values could still have high
significance. As a matter of fact, all values that we found, for all
types of scaling, indicate a significant discrepancy between the
corresponding distributions. However, we stress that here we are
considering the whole profile of the distribution, from the lowest to
the highest value of the performance variable. The most interesting
part of the distributions, and the one which is likely to reflect
collective social dynamics, is certainly the tail, because it is where
one has the largest cascades of votes for the same individual. On the contrary, the
initial part of the curve corresponds to poorly voted candidates, and
there are many ways to get to such modest outcomes (like being voted
solely by closest family members and friends), hardly
susceptible of a mathematical modelling. 
But at this stage we did not want
to identify the most ``interesting'' part of the distribution by
constraining the range of the variable, which is always
tricky. Therefore we decided to compare the full distributions.

We finally remark that in social dynamics one can hardly get the same
striking data collapses obtained in physical systems and models. Even if the
social atom hypothesis implies that just a few features of the social
actors and their interactions determine the large-scale behavior, the
complexity of human nature and context-dependent factors may still have
some influence, albeit small. For instance, in the Polish distributions of Fig. 1B
there is a hump for $v/v_0\approx 5$, which occurs systematically at
the national level, but which is absent in the other distributions of
the same class. Therefore, obtaining the agreement of the distributions shown in
Fig. 1F, despite
all differences between countries and historical ages, is truly remarkable.

\section{Discussion}
We have performed an empirical analysis of elections held
in 15 countries in various years. We focused on the competition
between candidates, which is a truly open competition when the voters can
indicate their favourite representatives in the ballot and
candidates with the largest number of votes are ranked the highest. This occurs
in proportional elections with open lists. Of the countries
for which we found data, 10 adopt open lists. Five of them (Group U), Italy, Finland, Poland,
Denmark and Estonia (since 2002) have very similar election rules, the
other five are characterized by important differences (e.g. compulsory
vote, huge districts and weak role of parties in Brazil), which are likely to affect the behavior
of voters and candidates, leading to measurable differences in the
statistical properties of the electoral outcomes. Indeed, the
distribution of the number of votes received by a candidate,
normalized by the average number of votes gained by his/her
competitors in the same party list, seems to be the same for the
nations of Group U, while there are marked differences from the curves
obtained from the other countries. This result, originally found by
Fortunato and Castellano for Italy, Finland and
Poland~\cite{fortunato07}, is confirmed here on a much larger data
collections and holds for Denmark and Estonia as well.

Different patterns are found for countries adopting semi-open lists,
in which in principle voters can choose the candidates, but the main
ranking criterion is still imposed by their party, regardless of the final
electoral score of the candidate, unless it exceeds a given threshold.
In this system the competition among the candidates is therefore not
really open, and it is no wonder that the distribution of electoral
performance does not follow the profile of the curves of Group U. 

In general we found that the shape of the distribution is much more sensitive to
the specific election rules adopted in the countries than to the
historical and cultural context where the election took place. This is evident when one
considers the evolution in time of distributions of any given country, which remain
essentially identical even after many years, if the voting system does not
change, but display visible
variations following the introduction and/or modification of election
rules as it happened in Estonia in 2002, Slovakia in 1994, Czech Republic in
2006. The case of Estonia is spectacular: before 2002 it used
semi-open lists, and the distributions of relative performance of a
candidate with respect to his/her party competitors did not compare
well with the curves of the other countries of Group U. After
the introduction of open lists, instead, the distributions became very
similar to the universal curve. 
Such sensitivity of the distributions might allow to detect anomalies, 
e.g. large-scale fraud, in future elections~\cite{baez06,klimek12}.

Our analysis proves that the success of a candidate, measured by the
number of votes, strongly depends on the party he/she belongs to, and
that only when one considers the competition among candidates of the
same party universal signatures may emerge. Indeed, neglecting the
party affiliation does not seem to take us very far: the two
party-independent normalizations we have considered, following the
procedure by Costa Filho et al.~\cite{costafilho99,costafilho03,
  araripe09}, do not seem to reveal strong common features
among distributions of different countries, not even when the latter
follow nearly identical election schemes (e.g. the nations of Group U).

\section{Methods}
\subsection{Election data}
Here we consider the data sets for parliamentary elections from $15$ countries
with open and semi-open lists: Italy (1958, 1972, 1976, 1979 and 1987) ~\cite{ItalyData}, Poland
(2001, 2005, 2007 and 2011) ~\cite{PolandData}, Finland (1995, 1999,
2003 and 2007) ~\cite{FinlandData},
Denmark (1990, 1994, 1998, 2001, 2005, 2007 and 2011) ~\cite{DenmarkData}, Estonia
(1992, 1995, 1999, 2003, 2007 and 2011) ~\cite{EstoniaData},
Slovenia (2004, 2008 and 2011) ~\cite{SloveniaData},
Greece (2007 and 2009) ~\cite{GreeceData},
Switzerland (2007 and 2011) ~\cite{SwissData},
Brazil (elections for state deputies in 2002, 2006 and 2010) ~\cite{BrazilData},
Uruguay (2004 and 2009) ~\cite{UruguayData},
Sweden (2006 and 2010) ~\cite{SwedenData},
Belgium (2007 and 2010) ~\cite{BelgiumData},
Slovakia (1994, 1998, 2002, 2010 and 2012) ~\cite{SlovakiaData1,SlovakiaData2},
Czech Republic (2002, 2006 and 2010) ~\cite{CzechData} and the Netherlands (2010 and 2012)~\cite{NetherlandsData}.
Further details and sources for each file are given in
Table~\ref{tab:datasrc} in Appendix, 
while the compiled and cleaned data maybe be downloaded at \url{http://becs.aalto.fi/en/research/complex_systems/elections/}.

\subsection{Comparing distributions}
We use the Kolmogorov-Smirnov (K-S) distance~\cite{NR} to measure the dissimilarity of two
empirical distributions. The K-S distance $D$ is defined as the maximum value
of the absolute difference between the corresponding cumulative distribution functions, i.e.
\begin{equation}
 D =\max_{x}|S_{N_{1}}(x)-S_{N_{2}}(x)|
 \label{ks_eq}
\end{equation}
where $S_{N_{1}}(x)$ and $S_{N_{2}}(x)$ are the cumulative distributions for two
data sets of size $N_{1}$ and $N_{2}$. 

Since we have multiple datasets for each country, in order to compute the dissimilarity of the distributions at the
national level and across countries we proceed as follows. For a given country $X$ we compute the 
distance between any two distributions for elections
of $X$. For a pair of countries $X$ and $Y$
we compute the distance between any pair of distributions $P_X$ and
$P_Y$, corresponding to one dataset of $X$ and one of $Y$,
respectively. In both cases we take the average $D_{avg}$ and the maximum
$D_{max}$ of the resulting values. In this way we estimate the average
and the maximum distance between distributions of the same country and
between distributions of two different countries.

\begin{acknowledgments}
We thank Lauri Loiskekoski for helping us to collect the election data.
We also thank Claudio Castellano  and Raimundo N. Costa Filho for useful
comments on the manuscript.
\end{acknowledgments}

\onecolumngrid
\newpage
\appendix

\numberwithin{table}{section}
\numberwithin{figure}{section}

%\section*{Supplementary material}

\section{Description of election systems}
\label{app1}

\paragraph{Italy}
During 1948 to 1992, the members of the Chamber of Deputies 
(\textit{Camera dei Deputati}) 
were elected by proportional representation (PR) 
in multi-member electoral districts, except in Valle d'Aosta where one member
was elected by simple majority.
Over this legislative period, Italy used an open-list PR system in which voters
could decide to use as many 
as 3 (4 for very large districts) preference votes for individual candidates on
the party list of their choice.
However, in 1992 this number was limited to unity.
In each constituency, seats were divided between open lists using the largest
remainder method with the 
Imperiali quota, and the remaining votes and seats were transferred to the
national level, 
where special closed lists of national leaders received the last seats using the
Hare quota.

\paragraph{Poland}
The lower chamber of the  Polish parliament, the \textit{Sejm}, has $460$ seats 
out of which $391$ are elected in the
multi-member districts and $69$ at the national level. The seats within the
district are allocated to a party or independent list according to d'Hondt
method, and reallocation to the candidates on each list is done according to
plurality rule. 

\paragraph{Finland}
The Finnish parliament, \textit{Eduskunta}, has $200$ seats, distributed among
$15$
multi-member districts. The candidates are nominated by a political party. A
political party presents a
list for each district, with at least $14$ candidates. In Finland there is no
electoral threshold and all seats are allocated within the electoral
constituencies. The voter is presented with the ballots from each party, and
he/she cast a vote for one candidate only, expressing this way a preference
toward a certain candidate, but also towards a certain party. The allocation of
the seats is according to d'Hondt constituency list system.

\paragraph{Denmark}
The parliament of the Kingdom of Denmark, the \textit{Folketing}, is composed of
$179$
members directly elected by a two-tier, six-stage proportional representation
system. $135$ of  the total $175$ members that represent metropolitan
Denmark are chosen in multi-member
electoral districts grouped into three electoral regions, while the remaining
$40$ seats are allocated to ensure proportionality at a national level. Voters
may cast a ballot for a district party list, or for a specific candidate. From
$1970$ to
$2005$ the seats within the constituencies were distributed according to the
modified
Sainte-Lagu\"{e} method of PR, while since $2007$ the seats are distributed
according to the d'Hondt or
largest average method of PR. The seats on the national level are apportioned
among political parties that obtain at least one district seat,  or
obtained as many votes as on average were cast per constituency seat in at least
two of the three regions, or at least two percent of all valid votes cast at the
national level. The total number of seats to which each cartel is
entitled is determined using the d'Hondt method. From this total number is then
subtracted
the number of constituency seats won by associated lists within each district.
The
difference gives the number of the forty supplemental seats to which the party
is entitled. Seats awarded on the national level are reallocated to each party's
component constituency lists by a two-step procedure. Seats are first allocated
to regions, by the Sainte-Lagu\"{e} method. Then, within regions, they are
allocated
to constituencies by another divisor method. These seats are then re reallocated
to each list's candidates, by three different procedures.

\paragraph{Estonia}
We consider the data from the elections to the Estonian parliament 
(\textit{Riigikogu}) for the period of two
decades ($1992$-$2012$) during which there were two reforms in the electoral
system in $1994$ and $2002$ \cite{renwick2011}. Keeping in mind the years of
these reforms, the data sets for Estonian elections can be categorized into 
three groups:
$1992$ as the first group, the second consists of the data from the elections held
in $1995$ and $1999$, and in the third  we consider elections after the
second reform, that of $2003$. 
The rules used in elections in $1992$ were partly
similar to those used in Finland \cite{grofman1999}. 
The country was divided into $12$ multi-member districts whose magnitudes ranged from
five to thirteen seats, whereas the whole Estonian parliament consists of $101$ members.
Each party presented a list of candidates and voters voted for an individual
candidate. Candidates who received a Hare quota were certified as
\textit{personally elected}. The remainder of votes were added by the list, and
if full quota materialized, the top voted candidates on the list received
district seats. Unlike Finland, where the seats are allocated in the district,
the remained unallocated seats were compiled nationwide and appointed to closed
lists of parties that received at least five percent of the votes on national
level. For the allocation on the national level the quasi-d'Hondt quota was used
\cite{grofman1999}. This kind of system led to selection of candidate which had
only $51$ personal votes. For this reasons, a restriction for
district seats was introduced for the elections held in $1995$: in order to be
chosen, a candidate had to win at least the number of votes equal to $0.1$ of
quota. The reform in $2002$ was related to lists at the national level, the
ordering of the candidates was according to the number of their personal votes. 
In this sense the later reform, the transition from semi-open to open lists led to
a greater personalization of the electoral system in Estonia~\cite{renwick2011}.

\paragraph{Slovenia}
In Slovenia, the deputies of the National Assembly (\textit{Dr\v{z}avni zbor}),
with the exception of the
two representatives of minorities, are elected by proportional representation,
with a $4\%$ electoral threshold required at the national level. The
country is divided into eight territorial constituencies, each represented by
eleven elected deputies. For the elections of the representatives of the Italian
and Hungarian ethnic communities, two special constituencies are formed, one for
each minority. The deputies representing the minorities are elected on the basis
of the majority principle. The seat allocation within the districts is as
follows: each list gets as many seats as there are whole Hare quotas contained
in its vote within the district. Seats unallocated within the districts are
aggregated at the national level and distributed by PR-d'Hondt rule, on the basis of
each party remainder vote (the sum of all remainders from associated
constituency lists). Only lists which won at least three seats are eligible to
participate in this step of seat allocation. The party seats won on the national
level are reallocated to the lists according to their ranking. The lists within
the party are ranked according to their remainder is expressed as
a fraction of the quota in its constituency. The lists from constituencies all
of whose seats have already been allocated are not considered in the
apportionment of national seats. The seats awarded to the lists are reallocated
to each list's candidates as follows and candidates on each list are associated 
with one (or two) geographically defined sub-districts. 
The candidates on each list are ranked in
terms of the percentage of the total vote each has received in his or her
sub-district. The top candidates on the list get the seats to which their list
is entitled.

\paragraph{Greece}
The Hellenic Parliamnet is composed of $300$ deputies elected for four-year term
through \textit{reinforced} proportional representation system. Greece is
divided into $56$ districts, out of which $48$ have more than one representatives 
in the Parliament. The winning party on the national level in Greece receives a
majority bonus of $40$  in $2007$ ($50$ in $2009$), while the remaining $260$ 
($250$ in $2009$) seats are distributed by the largest remainder
method (Hagenbach-Bischoff) of proportional representation (PR) on a nationwide
basis among parties polling at least $3\%$ of the vote. The voters can express their
preference toward certain candidates on the the party list. The number of
preference votes depends on the number of seats in constituency. Single-member
seats were filled by the plurality or first-past-the-post method,
in which the candidate obtaining the largest number of votes in the constituency
was elected to office. In multimember districts the seats won by party are
reallocated to each list's candidates by plurality. The only exception are the
party heads and acting or past Prime Ministers who are automatically placed at
the top of their party list.

\paragraph{Switzerland}
The National Council (\textit{Nationalrat/Conseil National/Consiglio
Nazionale/Cussegl Naziunal})
is composed of $200$ members
elected for a $4$-year term of office in $26$ constituencies - the
cantons of Switzerland. The electors have as many votes as there are allocated
seats for the district. These votes can be given to all candidates on a single
list or to candidates from different lists. The seats apportionment within the
constituency is based on Hagenbach-Bischoff method, while the list seats are
assigned to the candidates with the largest vote totals within each list. Like
in Poland, Finland, Denmark and Estonia, the voters have total control on who
will represent them in parliament.

\paragraph{Brazil}
In Brazil, the elections for the $513$ seats of the Chamber of Deputies 
(\textit{C\^{a}mara dos Deputados}) of the National Congress are conducted 
in $27$ multi-member ($8$ to $70$ seats, based on population) constituencies
corresponding to the country's $26$ states and the Federal district, using the 
party-list proportional system with seats allotted according to the simple
quotient and highest average calculations. 
The seats won by each list are in turn awarded to the
candidates on the basis of preferential votes cast by the electorate. 
Vacancies arising between general elections are filled by substitutes elected
at the same time as titular members. 
If no substitute is available and there remain at least 15 months before the
end of the term of the member concerned, by-elections are held. 
Furthermore, voting is compulsory, and abstention being punishable by a fine.

\paragraph{Uruguay}
The Chamber of Deputies (\textit{C\'{a}mara de Diputados}) in Uruguay has 
$99$ seats allocated in $19$ multi-member
districts through a list proportional representation system. The members
of the Chamber are elected for $5$ year term. 
Voting is compulsory in Uruguay and unjustified abstention is penalized by a fine.
The elections for Chamber are in
the same time as the elections for the Senate and presidential elections. For each
district the party can present more then one list, so called \textit{sub-lemas} 
containing usually a pair of candidates or more.
Uruguay uses the rule of the double simultaneous vote (DSV), which means that
the voter must vote for one of the \textit{sub-lemas} of the party he/she has
chosen in a Presidential contest. The distribution of seats between political
parties or cartels is decided by tallying the votes of each \textit{sub-lemas}.
The allocation of the seats is done according to variation of d'Hondt formula
devised by Borelli. The possible remaining seats are distributed on the national
level. Although the voters can not cast a vote for a single member but for the
\textit{sub-lemas}, we can use the quantity $v/v_{0}$ for measuring performance
of the \textit{sub-lemas} since the parties propose several \textit{sub-lemas}
which usually contain two or three names. The data we consider here, have
information about the number of votes cast for each of the lists. The
performance of the \textit{sub-lema} is expressed as the relation between the
number of votes won by \textit{sub-lema} and number of votes won in average by
all \textit{sub-lemas} proposed by a party or cartel in the district.

\paragraph{Sweden}
The elections to the Swedish parliament (\textit{Riksdag})
uses open lists system, but unlike in countries previously
discussed, the voters can choose between three
different types of ballots papers: the party ballot paper which has
simply the name of the party, the name ballot paper has a party name
followed by a list of candidates and alternatively, a voter can take
blank ballot paper and write a party name on it. Voters in Sweden can
either vote for a party without expressing a preference towards any
candidate, or can vote for a person on the list, thereby giving the
voice to one candidate and indirectly to his/her party. Seats are
allocated among the Swedish political parties proportionally using a
modified form of the Saint-Lagu\"{e} method. The candidates
from the each party are determined according by two factors: the
candidate's ranking by their party and the number of preference votes
the obtained in the elections. If the candidate receives a number of
personal votes equal or greater than $8\%$ of the party's
total amount of votes, he/she will be automatically shifted to the top of
the list, regardless of the previous ranking. This type of system
gives the voters a degree of power in choosing candidates from the
list, but not as much as in the case of Poland. This is reflected in
the way people cast their votes and how the performance of the
candidates are distributed. 

\paragraph{Belgium}
In Belgium, the Chamber (\textit{Kamer van Volksvertegenwoordigers/Chambre des
Représentants/Abgeordnetenkammer})
seats are filled in eleven constituencies.
The political parties present the lists of candidates and voters can either
indicate a preference for one or more of them, or vote for a party. The seats
are distributed in each district among the lists that receive at least of all
valid votes cast in the constituency according to d'Hondt system. The ordering
of the candidates on the lists is similar to Sweden, i.e. it is a combination
of party ordering and nominative votes.

\paragraph{Slovakia}
For Slovakia we consider several election years $1994$, $1998$, $2002$, $2010$
and $2012$ during which there were three reforms of electoral system, $1998$,
$1999$ and $2004$~\cite{renwick2011}. 
The National council (\textit{N\'{a}rodn\'{a} rada})
of the Slovakian parliament consists of
$150$ members chosen for four year term on proportional representation
elections. For the elections in $1994$ the country was divided into four
multi-member constituencies. Each political party was nominating their
candidates, by submitting the list of not more than $50$ candidates in each
constituency. The voters were allowed to vote for four candidates on the list of
the same party, expressing this way their preference toward certain
candidates. The seat allocation among the parties within the constituency
was according to the Hagenbach-Bischoff system. All parties with more than $5\%$ of
the votes on national level took part in the seat distribution. Within
individual political parties the mandates were distributed among candidates
nominated by these parties in the order of priority of the list of candidates.
If some of the candidates gained at least $10\%$ of
preferential votes of the total number of votes cast for a political party
within a constituency, he/she had an advantage over other candidates on the list,
regardless of his/hers previously determined order. If the political party had
the right to several mandates, and several candidates met the conditions
specified in the previous sentence, then the candidates obtained mandates in the
order which was determined by the largest number of preferential votes, cast for
them. In case of equal preferential votes, the position of candidate for Deputy 
in the list of candidates was decisive. 
After the elections in $1994$, Slovakia became one,
nationwide electoral constituency, where political parties or
coalitions could submit lists of $150$ candidates. This semi-open list
system puts elections in Slovakia between
elections in Netherlands, where the voters do not have an influence on the 
seat allocation within the party, and countries like Finland and
Poland, where the ordering of the candidates solely depends on the
preference of the voters.

\paragraph{Czech Republic}
The lower house of the Czech Republic, the Chamber of deputies
(\textit{Poslaneck\'{a} sn\v{e}movna})
is composed of $200$ members directly elected by proportional elections.
The apportionment of Chamber seats in each of the fourteen multi-member
districts among competing lists is done using d'Hondt rule. In order to
participate in the distribution of constituency seats, a party must
obtain at least $5\%$ of all valid votes cast at the national level,
while coalitions of $2$, $3$ and $4$ or more parties are required to obtain
at least $10$, $15$ and $20$ percent of the vote (previously $7$, $9$ and
$11$ percent) respectively. 
The electoral reforms of 2000 changed the number of preferences to $2$ and 
the seats within the list are allocated to
candidates in the order in which they appear on the list, but the candidates
which received at least $7\%$ (reduced from $10\%$) of the votes cast for their
party list have
priority in the allocation of seats, regardless of their
position on the list.  This was followed in the elections of 2002 and 2006.
However in 2010, the preferences were set to $4$ and the amount of necessary
preferences for the relevant candidate was reduced to $5\%$.

\paragraph{Netherlands}
The  House of Representatives 
(\textit{Tweede Kamer})
has a particular way of allocating  seats. Each
party can represent the list of $50$ candidates ($80$ if the party had
more than $15$ seats in the previous term) with \textit{predetermined} ordering
of the candidates. The first person on the list, 
\textit{list puller}, is usually appointed by the party to lead its
election campaign and is a candidate for the \textit{Prime Minister}. 
Although, parties may choose to compete with the
different candidates in each district, the seat allocation on the
national level results in nationwide lists. Since, the voters cannot
influence the ordering of the candidates, i.e. the lists are of closed
types, they often vote for the list puller and first several
candidates on the list, as the more popular and
appreciated members of the party. 

%This is known as
%\textit{rich-gets-richer} effect which is characterized with
%power-law behavior of the distribution of the relevant 
%quantities~\cite{Eggenberger1923mattew,Simon1955,Merton1968matthew,Price1976,
%Albert2002}. 

\newpage
\section{CAAMn scaling}
\label{app2}

%%%%%%%%%%%%%%%%%%%%%%%%%%%%%%%%%%%
\begin{figure}[ht]
\includegraphics[width=0.75\textwidth]{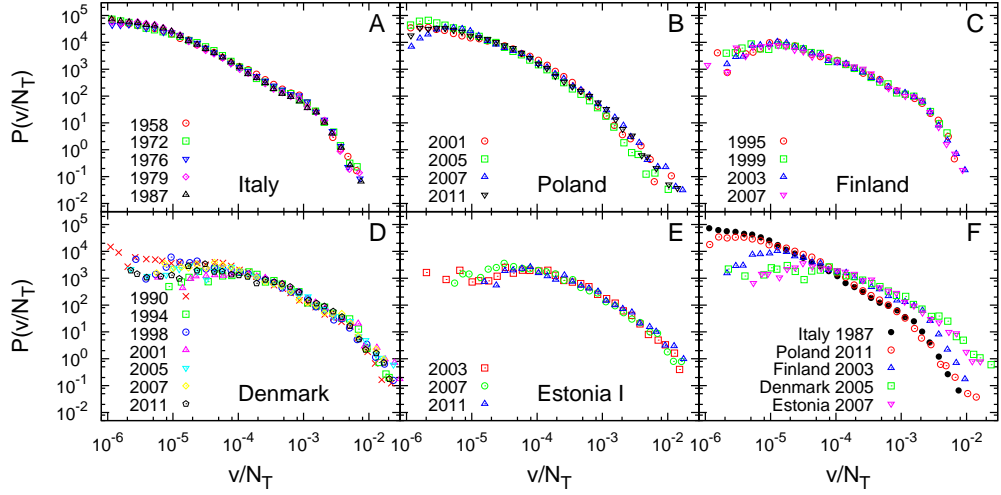}
\caption{Distributions of the fraction of votes obtained
by a candidate within the entire country
for different election years for (A)
Italy, (B) Poland, (C) Finland, (D) Denmark, (E) Estonia ($2003, 2007, 2011$). 
A comparison of the scaling curves for these countries is shown in (F).}
\label{figS1}
\end{figure}
%%%%%%%%%%%%%%%%%%%%%%%%%%%%%%%%%%%
%%%%%%%%%%%%%%%%%%%%%%%%%%%%%%%%%%%
\begin{figure}[ht]
\includegraphics[width=0.75\textwidth]{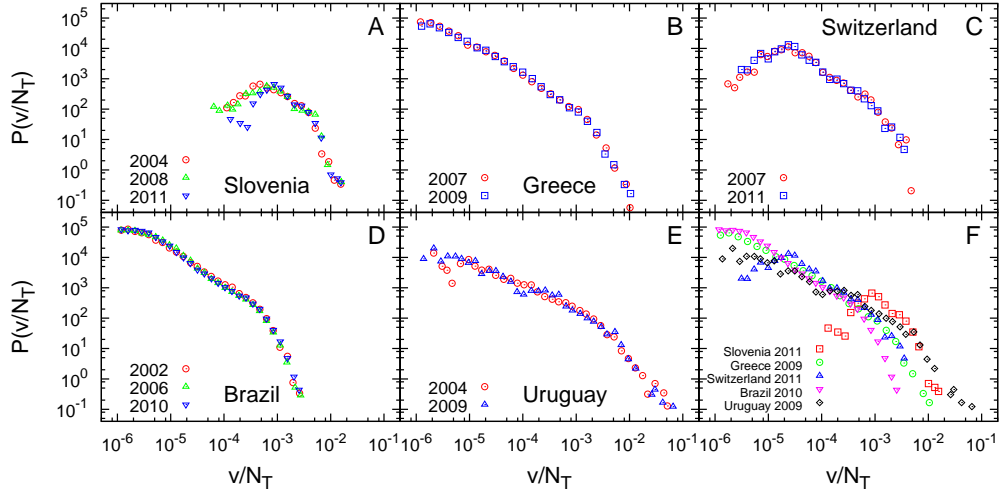}
\caption{Distributions of the fraction of votes obtained by
a candidate within the whole country 
for different election years for
Slovenia, ( Greece, (C) Switzerland, (D) Brazil and (E) Uruguay. 
A comparison of the scaling curves for these countries is shown in (F).}
\label{figS2}
\end{figure}
%%%%%%%%%%%%%%%%%%%%%%%%%%%%%%%%%%%
%%%%%%%%%%%%%%%%%%%%%%%%%%%%%%%%%%%
\begin{figure}[ht]
\includegraphics[width=0.75\textwidth]{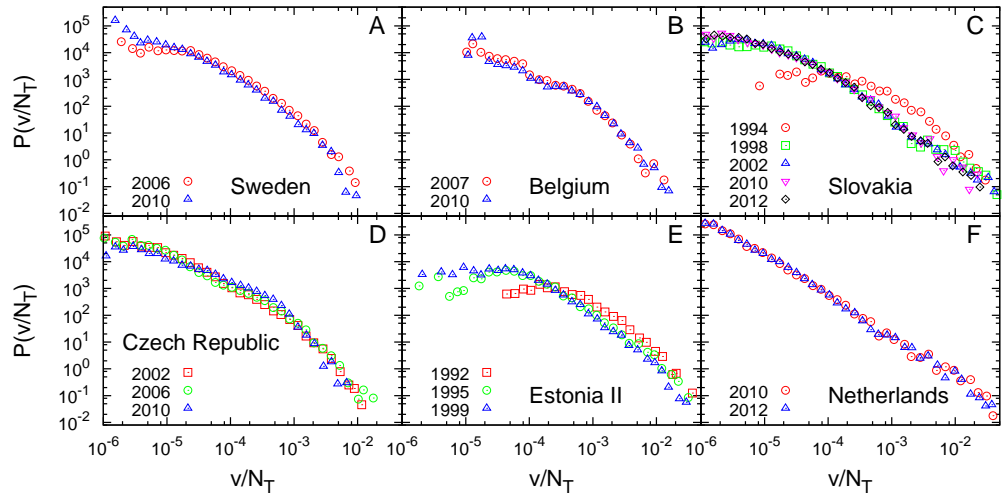}
\caption{Distributions of the fraction of votes obtained by
a candidate within the whole country 
for different election years in (A)
Sweden, (B) Belgium, (C) Slovakia, (D) Czech Republic, (E) Estonia (1992, 1995, 1999),
and (F) Netherlands.}
\label{figS3}
\end{figure}
%%%%%%%%%%%%%%%%%%%%%%%%%%%%%%%%%%%

\clearpage
\section{Tables}
\label{app3}

\begin{table}[ht]
 \caption{The estimates of $\gamma$ exponent for CAAMd scaling for open-list electoral systems.
 For Poland and Uruguay, it is not possible to find any finite range of values for which a power-law could be fitted.}
\begin{tabular}{|l|l|l|l|l|}
 \hline
Country & Year & $x_{min}$ & $x_{max}$ & $\gamma$ \\
\hline
Italy & 1958 & $0.00021$ & $0.04871$ & $-1.27(4)$ \\
\hline
Poland &  & n.d. & n.d. & n.d.\\
\hline
Finland & 1999 & $0.00036$ & $0.02879$ & $-1.11(3)$ \\
\hline
Denmark & 2007 & $0.00063$ & $0.02959$ & $-1.08(3)$ \\
\hline
Estonia & 2011 & $0.00330$ & $0.08730$ & $-1.47(5)$ \\
\hline
Greece & 2007 & $0.00008$ & $0.03403$ & $-1.10(4)$ \\
\hline
Switzerland & 2007 & $0.00063$ & $0.02795$ & $-1.21(4)$ \\
\hline 
Brazil & 2006 & $0.00007$ & $0.00968$ & $-1.12(1)$ \\
\hline
Uruguay &  & n.d. & n.d. & n.d.\\
\hline
\end{tabular}
\label{tab:exp}
\end{table}

\begin{table}[ht]\scriptsize
\caption{FC scaling: Average distance $D$ between datasets from different countries.}
\begin{tabular}{|l||l|l|l|l|l|l|l|l|l|l|l|l|l|l|l|l|}
\hline
 & It & Fi & Pl & Dk & EeI & EeII & Si & Gr & Ch & Br & Uy & Se & Be & Sk & Cz & Nl  \\
\hline
It & 0.03801 &0.06560 &0.05380 &0.09199 &0.07412 &0.15788 &0.37039 &0.29121 &0.45663 &0.23223 &0.14801 &0.19722 &0.28724 &0.27584 &0.12724 &0.65437 \\
\hline
Fi & &0.02592 &0.10756 &0.06669 &0.11274 &0.16889 &0.31749 &0.23889 &0.40258 &0.25590 &0.14209 &0.23261 &0.23913 &0.30750 &0.07937 &0.67421 \\
\hline
Pl & & &0.04236 &0.12456 &0.07161 &0.15240 &0.41216 &0.33374 &0.49942 &0.21605 &0.18217 &0.17276 &0.32289 &0.25497 &0.16750 &0.64148 \\
\hline
Dk & & & &0.08410 &0.12214 &0.16958 &0.31303 &0.23599 &0.39847 &0.23272 &0.13046 &0.23501 &0.23729 &0.29784 &0.09915 &0.64581 \\
\hline
EeI & & & & &0.05032 &0.11546 &0.40584 &0.32818 &0.48532 &0.20248 &0.13599 &0.14914 &0.34322 &0.22991 &0.17743 &0.62883 \\
\hline
EeII & & & & & &0.14210 &0.43881 &0.36203 &0.51480 &0.14974 &0.16591 &0.11289 &0.37911 &0.19494 &0.21826 &0.56205 \\
\hline
Si & & & & & & &0.06288 &0.08608 &0.11806 &0.46576 &0.33138 &0.52184 &0.18294 &0.56076 &0.25497 &0.82169 \\
\hline
Gr & & & & & & & &0.02087 &0.17273 &0.39836 &0.26706 &0.44342 &0.14608 &0.48803 &0.18109 &0.77310 \\
\hline
Ch & & & & & & & & &0.17259 &0.53076 &0.39489 &0.59665 &0.22172 &0.63361 &0.34065 &0.86106 \\
\hline
Br & & & & & & & & & &0.08174 &0.16535 &0.12180 &0.42645 &0.16871 &0.28469 &0.43365 \\
\hline
Uy & & & & & & & & & & &0.11100 &0.23317 &0.29583 &0.27973 &0.16721 &0.55188 \\
\hline
Se & & & & & & & & & & & &0.07814 &0.45601 &0.14886 &0.29123 &0.49503 \\
\hline
Be & & & & & & & & & & & & &0.16144 &0.51014 &0.18404 &0.80793 \\
\hline
Sk & & & & & & & & & & & & & &0.18281 &0.35344 &0.45883 \\
\hline
Cz & & & & & & & & & & & & & & &0.11962 &0.69293 \\
\hline
Nl & & & & & & & & & & & & & & & &0.10844 \\
\hline
\end{tabular}
\label{tab:FCclassIa}
\end{table}

\begin{table}[ht]\scriptsize
\caption{FC scaling: Maximum distance $D$ between datasets from different countries.}
\begin{tabular}{|l||l|l|l|l|l|l|l|l|l|l|l|l|l|l|l|l|}
\hline
 & It & Fi & Pl & Dk & EeI & EeII & Si & Gr & Ch & Br & Uy & Se & Be & Sk & Cz & Nl \\
\hline
It & 0.07070 &0.09998 &0.10859 &0.16108 &0.10361 &0.21909 &0.43408 &0.31982 &0.54035 &0.31357 &0.20025 &0.24036 &0.33814 &0.40206 &0.26237 &0.69261 \\
\hline
Fi & &0.03241 &0.14086 &0.13231 &0.13737 &0.25200 &0.35904 &0.24493 &0.46759 &0.33103 &0.16655 &0.27059 &0.26516 &0.42020 &0.19063 &0.71052 \\
\hline
Pl & & &0.05897 &0.19237 &0.08694 &0.19499 &0.47491 &0.36068 &0.57986 &0.29779 &0.24170 &0.22232 &0.37147 &0.38301 &0.29832 &0.67978 \\
\hline
Dk & & & &0.18642 &0.20043 &0.30713 &0.45101 &0.33843 &0.54346 &0.34308 &0.18052 &0.32110 &0.36328 &0.44594 &0.28313 &0.69737 \\
\hline
EeI & & & & &0.07077 &0.18239 &0.45052 &0.34142 &0.54968 &0.28357 &0.18697 &0.20743 &0.37557 &0.36842 &0.29624 &0.67421 \\
\hline
EeII & & & & & &0.20064 &0.55498 &0.44567 &0.64241 &0.27050 &0.26773 &0.21713 &0.48052 &0.36279 &0.40379 &0.64199 \\
\hline
Si & & & & & & &0.06963 &0.12116 &0.18554 &0.55632 &0.38073 &0.57478 &0.20762 &0.66732 &0.34597 &0.85289 \\
\hline
Gr & & & & & & & &0.02087 &0.23365 &0.46823 &0.27242 &0.46320 &0.16256 &0.57763 &0.23186 &0.79092 \\
\hline
Ch & & & & & & & & &0.17259 &0.62214 &0.45975 &0.65662 &0.34539 &0.73473 &0.45101 &0.88318 \\
\hline
Br & & & & & & & & & &0.11955 &0.23797 &0.16632 &0.51099 &0.22314 &0.43536 &0.50603 \\
\hline
Uy & & & & & & & & & & &0.11100 &0.27942 &0.31512 &0.35771 &0.23877 &0.59656 \\
\hline
Se & & & & & & & & & & & &0.07814 &0.49512 &0.23553 &0.41632 &0.56354 \\
\hline
Be & & & & & & & & & & & & &0.16144 &0.62021 &0.24679 &0.83190 \\
\hline
Sk & & & & & & & & & & & & & &0.30812 &0.54228 &0.58983 \\
\hline
Cz & & & & & & & & & & & & & & &0.17079 &0.77514 \\
\hline
Nl & & & & & & & & & & & & & & & &0.10844 \\
\hline
\end{tabular}
\label{tab:FCclassIm}
\end{table}

%%%%%%%%%%%%%%%%%%%%%%%%%%%%%%%%%%%%%%%%%%%%%%%%%%%%
\begin{table}[ht]\scriptsize
\caption{CAAMd scaling: Average distance $D$ between datasets from different countries.}
\begin{tabular}{|l||l|l|l|l|l|l|l|l|l|l|l|l|l|l|l|l|}
\hline
 & It & Fi & Pl & Dk & EeI & EeII & Si & Gr & Ch & Br & Uy & Se & Be & Sk & Cz & Nl \\
\hline
It & 0.06929 &0.24181 &0.19605 &0.43669 &0.39816 &0.37895 &0.71540 &0.16986 &0.20398 &0.15814 &0.49786 &0.19310 &0.17631 &0.51498 &0.13853 &0.82727 \\
\hline
Fi &  &0.04488 &0.09237 &0.22718 &0.17321 &0.18166 &0.54486 &0.12413 &0.07568 &0.34413 &0.38077 &0.20539 &0.13957 &0.59584 &0.31366 &0.90123 \\
\hline
Pl &  & &0.05296 &0.30310 &0.24112 &0.21381 &0.62367 &0.10381 &0.05648 &0.31814 &0.44676 &0.17102 &0.07638 &0.59226 &0.28284 &0.89917 \\
\hline
Dk &  & & &0.14625 &0.14076 &0.23600 &0.34091 &0.28756 &0.28417 &0.50810 &0.23662 &0.40870 &0.35072 &0.71065 &0.48959 &0.93467 \\
\hline
EeI &  & & & &0.16015 &0.20126 &0.41525 &0.26706 &0.23025 &0.49047 &0.31402 &0.35743 &0.29123 &0.70049 &0.46403 &0.94112 \\
\hline
EeII &  & & & & &0.27932 &0.49650 &0.26027 &0.20853 &0.48211 &0.36498 &0.32008 &0.25375 &0.69835 &0.45156 &0.94359 \\
\hline
Si &  & & & & & &0.12409 &0.55850 &0.59761 &0.75097 &0.27246 &0.70531 &0.66668 &0.87455 &0.74770 &0.98217 \\
\hline
Gr &  & & & & & & &0.02514 &0.08957 &0.23211 &0.37571 &0.18430 &0.12873 &0.53297 &0.22762 &0.85404 \\
\hline
Ch &  & & & & & & & &0.03784 &0.31057 &0.40198 &0.16720 &0.09645 &0.57467 &0.27950 &0.89779 \\
\hline
Br &  & & & & & & & & &0.04599 &0.53713 &0.20996 &0.31572 &0.42003 &0.09011 &0.77025 \\
\hline
Uy &  & & & & & & & & & &0.07334 &0.50697 &0.48601 &0.75089 &0.54306 &0.93769 \\
\hline
Se &  & & & & & & & & & & &0.31276 &0.19498 &0.51565 &0.20655 &0.82880 \\
\hline
Be &  & & & & & & & & & & & &0.07014 &0.60935 &0.27882 &0.92955 \\
\hline
Sk &  & & & & & & & & & & & & &0.37043 &0.45729 &0.64184 \\
\hline
Cz &  & & & & & & & & & & & & & &0.14945 &0.79005 \\
\hline
Nl &  & & & & & & & & & & & & & & &0.11057 \\
\hline
\end{tabular}
\label{tab:da}
\end{table}
%%%%%%%%%%%%%%%%%%%%%%%%%%%%%%%%%%%%%%%%%%%%%%%%%%%%
%%%%%%%%%%%%%%%%%%%%%%%%%%%%%%%%%%%%%%%%%%%%%%%%%%%%
\begin{table}[ht]\scriptsize
\caption{CAAMd scaling: Maximum distance $D$ between datasets from different countries.}
\begin{tabular}{|l||l|l|l|l|l|l|l|l|l|l|l|l|l|l|l|l|}
\hline
 & It & Fi & Pl & Dk & EeI & EeII & Si & Gr & Ch & Br & Uy & Se & Be & Sk & Cz & Nl \\
\hline
It & 0.09564 &0.31050 &0.27939 &0.59191 &0.51746 &0.56189 &0.79147 &0.21808 &0.25947 &0.20313 &0.54676 &0.29096 &0.24784 &0.59231 &0.19929 &0.83707 \\
\hline
Fi &  &0.06399 &0.12927 &0.37020 &0.25594 &0.31480 &0.60435 &0.14990 &0.10232 &0.38160 &0.40233 &0.32485 &0.17915 &0.75149 &0.40082 &0.91477 \\
\hline
Pl &  & &0.07079 &0.46116 &0.33911 &0.40269 &0.69504 &0.11828 &0.07156 &0.34448 &0.46076 &0.29170 &0.13499 &0.72876 &0.36182 &0.91553 \\
\hline
Dk &  & & &0.27609 &0.28350 &0.46215 &0.52053 &0.38945 &0.41921 &0.61449 &0.31473 &0.61296 &0.50386 &0.89750 &0.65247 &0.95876 \\
\hline
EeI &  & & & &0.16015 &0.32867 &0.52946 &0.32185 &0.31209 &0.55802 &0.33626 &0.53202 &0.39225 &0.87834 &0.58936 &0.95553 \\
\hline
EeII &  & & & & &0.40882 &0.70318 &0.35761 &0.36411 &0.59422 &0.47135 &0.57410 &0.44274 &0.90111 &0.62867 &0.96391 \\
\hline
Si &  & & & & & &0.15834 &0.60628 &0.65331 &0.79363 &0.30116 &0.81381 &0.72674 &0.97543 &0.83624 &0.98663 \\
\hline
Gr &  & & & & & & &0.02514 &0.09492 &0.24590 &0.38503 &0.23799 &0.13999 &0.62957 &0.28286 &0.86564 \\
\hline
Ch &  & & & & & & & &0.03784 &0.32906 &0.41469 &0.27331 &0.10927 &0.69802 &0.34844 &0.92066 \\
\hline
Br &  & & & & & & & & &0.04599 &0.55228 &0.37209 &0.36614 &0.46050 &0.18873 &0.78936 \\
\hline
Uy &  & & & & & & & & & &0.07334 &0.56762 &0.49001 &0.86231 &0.59356 &0.94393 \\
\hline
Se &  & & & & & & & & & & &0.31276 &0.30932 &0.74055 &0.38563 &0.87916 \\
\hline
Be &  & & & & & & & & & & & &0.07014 &0.75194 &0.36306 &0.93611 \\
\hline
Sk &  & & & & & & & & & & & & &0.82481 &0.56383 &0.93643 \\
\hline
Cz &  & & & & & & & & & & & & & &0.20000 &0.82044 \\
\hline
Nl &  & & & & & & & & & & & & & & &0.11057 \\
\hline
\end{tabular}
\label{tab:dm}
\end{table}
%%%%%%%%%%%%%%%%%%%%%%%%%%%%%%%%%%%%%%%%%%%%%%%%%%%%

%%%%%%%%%%%%%%%%%%%%%%%%%%%%%%%%%%%%%%%%%%%%%%%%%%%%
\begin{table}[ht]\scriptsize
\caption{CAAMn scaling: Average distance $D$ between datasets from different countries.}
\begin{tabular}{|l||l|l|l|l|l|l|l|l|l|l|l|l|l|l|l|l|}
\hline
 & It & Fi & Pl & Dk & EeI & EeII & Si & Gr & Ch & Br & Uy & Se & Be & Sk & Cz & Nl \\
It & 0.09735 &0.40966 &0.14123 &0.58838 &0.61612 &0.58557 &0.85043 &0.11092 &0.41552 &0.11946 &0.51910 &0.20152 &0.39650 &0.25389 &0.13287 &0.45861 \\
\hline
Fi & &0.04351 &0.35011 &0.23342 &0.24521 &0.21334 &0.64307 &0.31078 &0.13590 &0.47827 &0.24283 &0.34281 &0.31376 &0.31712 &0.33150 &0.74166 \\
\hline
Pl & & &0.09321 &0.55354 &0.57673 &0.52816 &0.87547 &0.12517 &0.31186 &0.21890 &0.51622 &0.18065 &0.34107 &0.18941 &0.14757 &0.55727 \\
\hline
Dk & & & &0.12727 &0.10739 &0.23872 &0.45690 &0.49039 &0.34408 &0.62871 &0.13951 &0.54174 &0.45328 &0.44034 &0.50761 &0.82775 \\
\hline
EeI & & & & &0.08995 &0.23476 &0.47561 &0.51604 &0.36519 &0.65453 &0.14583 &0.56573 &0.47355 &0.45637 &0.53249 &0.85558 \\
\hline
EeII & & & & & &0.31559 &0.56396 &0.48834 &0.29685 &0.63564 &0.26306 &0.51877 &0.45609 &0.45362 &0.50424 &0.84092 \\
\hline
Si & & & & & & &0.12035 &0.78621 &0.71714 &0.88896 &0.42180 &0.86111 &0.75142 &0.75679 &0.81878 &0.95440 \\
\hline
Gr & & & & & & & &0.04070 &0.32655 &0.17752 &0.43189 &0.16606 &0.36930 &0.20312 &0.11543 &0.46402 \\
\hline
Ch & & & & & & & & &0.09460 &0.49885 &0.32350 &0.30518 &0.32525 &0.31319 &0.34490 &0.76336 \\
\hline
Br & & & & & & & & & &0.04866 &0.54868 &0.22568 &0.44326 &0.33406 &0.18699 &0.39566 \\
\hline
Uy & & & & & & & & & & &0.06401 &0.50106 &0.39542 &0.42237 &0.45469 &0.73689 \\
\hline
Se & & & & & & & & & & & &0.29447 &0.34396 &0.25998 &0.20101 &0.51794 \\
\hline
Be & & & & & & & & & & & & &0.39807 &0.36546 &0.36481 &0.59617 \\
\hline
Sk & & & & & & & & & & & & & &0.28602 &0.23325 &0.62914 \\
\hline
Cz & & & & & & & & & & & & & & &0.17667 &0.52661 \\
\hline
Nl & & & & & & & & & & & & & & & &0.11067 \\
\hline
\end{tabular}
\label{tab:na}
\end{table}
%%%%%%%%%%%%%%%%%%%%%%%%%%%%%%%%%%%%%%%%%%%%%%%%%%%%
%%%%%%%%%%%%%%%%%%%%%%%%%%%%%%%%%%%%%%%%%%%%%%%%%%%%
\begin{table}[ht]\scriptsize
\caption{ CAAMn scaling: Maximum distance $D$ between datasets from different countries.}
\begin{tabular}{|l||l|l|l|l|l|l|l|l|l|l|l|l|l|l|l|l|}
\hline
 & It & Fi & Pl & Dk & EeI & EeII & Si & Gr & Ch & Br & Uy & Se & Be & Sk & Cz & Nl \\
\hline
It & 0.16596 &0.50374 &0.26781 &0.71614 &0.69386 &0.77979 &0.90781 &0.18572 &0.49951 &0.19006 &0.57505 &0.35774 &0.48851 &0.71299 &0.30706 &0.52575 \\
\hline
Fi & &0.06925 &0.44934 &0.33086 &0.29380 &0.41589 &0.69992 &0.35526 &0.19113 &0.51624 &0.25757 &0.46646 &0.39807 &0.36838 &0.44178 &0.74807 \\
\hline
Pl & & &0.15795 &0.69344 &0.65574 &0.77212 &0.92970 &0.15241 &0.41221 &0.29297 &0.55847 &0.24358 &0.39807 &0.68071 &0.23487 &0.60646 \\
\hline
Dk & & & &0.25816 &0.22690 &0.38805 &0.63695 &0.57707 &0.46020 &0.71573 &0.19562 &0.69262 &0.60308 &0.62669 &0.66268 &0.87907 \\
\hline
EeI & & & & &0.12580 &0.35347 &0.57749 &0.55147 &0.40626 &0.69928 &0.17781 &0.66887 &0.55584 &0.57848 &0.64139 &0.87960 \\
\hline
EeII & & & & & &0.46472 &0.76296 &0.64484 &0.53917 &0.77335 &0.32213 &0.76560 &0.67811 &0.70570 &0.72553 &0.91809 \\
\hline
Si & & & & & & &0.15350 &0.82560 &0.78469 &0.92936 &0.50423 &0.91929 &0.83452 &0.89726 &0.88466 &0.96836 \\
\hline
Gr & & & & & & & &0.04070 &0.34150 &0.20807 &0.43647 &0.21221 &0.39807 &0.56949 &0.15117 &0.48829 \\
\hline
Ch & & & & & & & & &0.09460 &0.52232 &0.34134 &0.44541 &0.39807 &0.43729 &0.42843 &0.76747 \\
\hline
Br & & & & & & & & & &0.05984 &0.57857 &0.38059 &0.51411 &0.71360 &0.33074 &0.41760 \\
\hline
Uy & & & & & & & & & & &0.06401 &0.55696 &0.46762 &0.50409 &0.51926 &0.75746 \\
\hline
Se & & & & & & & & & & & &0.29447 &0.42572 &0.68662 &0.27888 &0.65408 \\
\hline
Be & & & & & & & & & & & & &0.39807 &0.58983 &0.39807 &0.79575 \\
\hline
Sk & & & & & & & & & & & & & &0.60654 &0.65949 &0.87480 \\
\hline
Cz & & & & & & & & & & & & & & &0.23837 &0.58946 \\
\hline
Nl & & & & & & & & & & & & & & & &0.11067 \\
\hline
\end{tabular}
\label{tab:nm}
\end{table}
%%%%%%%%%%%%%%%%%%%%%%%%%%%%%%%%%%%%%%%%%%%%%%%%%%%%

\newpage
%%%%%%%%%%%%%%%%%%%%%%%%%%%%%%%%%%%%%%%%%%%%%%%%%%%%%
%\subsection*{IV. DATA SOURCES}\label{sec:datasources}
%%%%%%%%%%%%%%%%%%%%%%%%%%%%%%%%%%%%%%%%%%%%%%%%%%%%%

 %%%%%%%%%%%%%%%%%%%%%%%%%%%%%%%%%%%%%%%%%%%%%%%%%%%%
\begin{table}[ht]\footnotesize
%\caption{Data details and sources}
\caption{Data for parliamentary elections for several years in 
 Italy~\cite{ItalyData}, Poland~\cite{PolandData}, Finland~\cite{FinlandData},
Denmark~\cite{DenmarkData}, Estonia~\cite{EstoniaData},
 Sweden~\cite{SwedenData}, Belgium~\cite{BelgiumData},
Switzerland~\cite{SwissData}, 
 Slovenia~\cite{SloveniaData}, Czech Republic~\cite{CzechData},
Greece~\cite{GreeceData}, 
 Slovakia~\cite{SlovakiaData1,SlovakiaData2},
Netherlands~\cite{NetherlandsData}, 
 Uruguay~\cite{UruguayData} and Brazil~\cite{BrazilData} were collected.
 Further details regarding the individual data files are given below.}
\begin{tabular}{||l|l|l|l|l||}
\hline
Country & Year  & Valid Votes & Voting \% &     Data source\\
\hline
Italy   &1958   & 30434681      & 93.83 &
\verb+http://elezionistorico.interno.it/index.php?tpel=C&dtel=25/05/1958+ \\
\hline
        &1972   & 34532535      & 93.21 &
\verb+http://elezionistorico.interno.it/index.php?tpel=C&dtel=07/05/1972+ \\
\hline
        &1976   & 37755090      & 93.39 &
\verb+http://elezionistorico.interno.it/index.php?tpel=C&dtel=20/06/1976+ \\
\hline
        &1979   & 38242918      & 90.62 &
\verb+http://elezionistorico.interno.it/index.php?tpel=C&dtel=03/06/1979+ \\
\hline
        &1987   & 40586573      & 88.83 &
\verb+http://elezionistorico.interno.it/index.php?tpel=C&dtel=14/06/1987+ \\
\hline
Poland  &2001   & 13559412      & 46.29 & \verb+http://wybory2011.pkw.gov.pl+ \\
\hline
        &2005   & 12244903      & 40.5  &
\verb+http://www.wybory2005.pkw.gov.pl+ \\
\hline
        &2007   & 16477734      & 53.9  & \verb+http://wybory2007.pkw.gov.pl+ \\
\hline
        &2011   & 15050027      & 48.92 & \verb+http://wybory2011.pkw.gov.pl+ \\
\hline
Finland &1995   & 2803602       & 68.6  &
\verb+http://tilastokeskus.fi/tk/he/vaalit/vaalit95/risvaa.xls+\\
\hline
        &1999   & 2710095       & 65.3  &
\verb+http://tilastokeskus.fi/tk/he/vaalit/vaalit99/ehdokas.xls+\\
\hline
        &2003   & 2815700       & 66.7  & {\scriptsize
\verb+http://tilastokeskus.fi/tk/he/vaalit/vaalit2003/vaalit2003_vaalitilastot_ehdokkaatverluku.xls+}\\
\hline
        &2007   & 2790752       & 65    &
\verb+http://www.stat.fi/til/evaa/2007/evaa_2007_2007-03-28_tau_015.xls+\\
\hline
Denmark &1990   & 3239468       & 82.19 & {\tiny
\verb+http://sum.dk/Aktuelt/Publikationer/Publikationer_IN/~/media/Filer-Publikationer-IN/Valg/1996/FTvalg-1990/FTvalg-1990.ashx+}\\
\hline
        &1994   & 3327597       & 83.42 & {\tiny
\verb+http://sum.dk/Aktuelt/Publikationer/Publikationer_IN/~/media/Filer-Publikationer-IN/Valg/1996/FTvalg-1994/FTvalg-1994.ashx+}\\
\hline
        &1998   & 3431926       & 85.94 & {\tiny
\verb+http://sum.dk/Aktuelt/Publikationer/Publikationer_IN/~/media/Filer-Publikationer-IN/Valg/1999/FTvalg-1998/FTvalg-1998.ashx+}\\
\hline
        &2001   & 3449668       & 86.26 & {\tiny
\verb+http://sum.dk/Aktuelt/Publikationer/Publikationer_IN/~/media/Filer-Publikationer-IN/Valg/2003/FTvalg-2001/FTvalg-2001.ashx+}\\
\hline
        &2005   & 3357212       & 83.85 & {\tiny
\verb+http://sum.dk/Tal-og-analyser/Valg/Folketingsvalg/~/media/Filer-Publikationer-IN/Valg/2006/FTV-2005.ashx+}\\
\hline
        & 2007  & 3459420       & 85.99 & {\tiny
\verb+http://www.statbank.dk/statbank5a/SelectVarVal/Define.asp?MainTable=FV07KAND&PLanguage=1&PXSId=0+}\\
\hline
        & 2011  & 3588919       & 87.2  & {\tiny
\verb+http://www.statbank.dk/statbank5a/SelectVarVal/Define.asp?MainTable=FV11KAND&PLanguage=1&PXSId=0+}\\
\hline
Estonia & 1992  & 467628        & 67.84 & \verb+http://vvk.ee/r92+\\
\hline
        & 1995  & 545825        & 69.06 & \verb+http://vvk.ee/r95+\\
\hline
        & 1999  & 492356        & 57.43 & \verb+http://vvk.ee/r99+\\
\hline
        & 2003  & 500686        & 58.2  & \verb+http://vvk.ee/r03+\\
\hline
        & 2007  & 555463        & 61.9  & \verb+http://vvk.ee/r07+\\
\hline
        & 2011  & 580264        & 63.5  &
\verb+http://www.vvk.ee/varasemad/rk2011+\\
\hline
Sweden  & 2006  & 5551278       & 81.99 &
\verb+http://www.val.se/val/val2006/slutlig/R/rike/roster.html+\\
\hline
        & 2010  & 5960408       & 84.63 &
\verb+http://www.val.se/val/val2010/slutresultat/R/rike/index.html+\\
\hline
Belgium & 2007  & 6671360       & 86.41 &
\verb+http://polling2007.belgium.be/en/index-2.html+\\
\hline
        & 2010  & 6527367       & 84.03 &
\verb+http://polling2010.belgium.be/en/index.html+\\
\hline
Switzerland     & 2007  & 2304090       & 46.87 & {\scriptsize
\verb+http://www.bfs.admin.ch/bfs/portal/fr/index/themen/17/02/blank/data/04/02.Document.110436.xls+}\\
\hline
        & 2011  & 2485803       & 48.5  & {\scriptsize
\verb+http://www.bfs.admin.ch/bfs/portal/fr/index/themen/17/02/blank/data/05/02.Document.155156.xls+}\\
\hline
Slovenia        & 2004  & 991263        & 60.65 &
\verb+http://volitve.gov.si/dz2004/en/index.htm+\\
\hline
        & 2008  & 1070523       & 63.1  &
\verb+http://volitve.gov.si/dz2008/en/index.html+\\
\hline
        & 2011  & 1121573       & 65.6  &
\verb+http://volitve.gov.si/dz2011/en/rezultati/izidi_enote.html+\\
\hline
Czech   & 2002  & 4768006       & 58    &
\verb+http://www.volby.cz/pls/ps2002/psm+\\
\hline
        & 2006  & 5348976       & 64.47 &
\verb+http://www.volby.cz/pls/ps2006/ps+\\
\hline
        & 2010  & 5230859       & 62.6  &
\verb+http://www.volby.cz/pls/ps2010/ps+\\
\hline
Greece  & 2007  & 7159006       & 72.18 &
\verb+http://ekloges-prev.singularlogic.eu/v2007/pages_en/index.html+\\
\hline
        & 2009  & 6858342       & 69.04 &
\verb+http://ekloges-prev.singularlogic.eu/v2009/pages/index.html?lang=en+\\
\hline
Slovakia        & 1994  & 2875458       & 74.18 &
\verb+http://www2.essex.ac.uk/elect/database/indexCountry.asp?country=Slovakia&opt=can+\\
\hline
        & 1998  & 3359176       & 83.5  &
\verb+http://www2.essex.ac.uk/elect/database/indexCountry.asp?country=Slovakia&opt=can+\\
\hline
        & 2002  & 2875081       & 69.15 &
\verb+http://www2.essex.ac.uk/elect/database/indexCountry.asp?country=Slovakia&opt=can+\\
\hline
        & 2010  & 2529385       & 58.83 &
\verb+http://app.statistics.sk/nrsr_2010+\\
\hline
        & 2012  & 2553726       & 59.11 &
\verb+http://app.statistics.sk/nrsr2012+\\
\hline
Netherlands     & 2010  & 9416001       & 75.18 &
\verb+http://www.verkiezingsuitslagen.nl+\\
\hline
        & 2012  & 9424235       & 74.57 &
\verb+http://www.verkiezingsuitslagen.nl+\\
\hline
Uruguay & 2004  & 2229583       & 89.62 &
\verb+http://www.corteelectoral.gub.uy/nacionales20041031+\\
\hline
        & 2009  & 2303336       & 89.86 &
\verb+http://elecciones.corteelectoral.gub.uy/20091025+\\
\hline
Brazil  & 2002  & 87474543      & 92.3  & {\scriptsize
\verb+http://www.tse.jus.br/eleicoes/eleicoes-anteriores/eleicoes-2002/resultado-da-eleicao-2002+}\\
\hline
        & 2006  & 93184830      & 88.9  & {\scriptsize
\verb+http://www.tse.jus.br/eleicoes/eleicoes-anteriores/eleicoes-2006/resultado-da-eleicao-2006+}\\
\hline
        & 2010  & 96580011      & 87    & {\scriptsize
\verb+http://www.tse.jus.br/eleicoes/eleicoes-anteriores/eleicoes-2010/estatisticas-de-candidaturas+}\\
\hline
\end{tabular}
\label{tab:datasrc}
\end{table}
%%%%%%%%%%%%%%%%%%%%%%%%%%%%%%%%%%%%%%%%%%%%%%%%%%%

\end{document}